\documentstyle[12pt,fleqn,newlfont]{article}
\newcommand{\half}{\frac{1}{2}}

\newcommand{\E}{\varepsilon}

\newcommand{\eps}{\epsilon}

\newcommand{\beq}{\begin{equation}}
\newcommand{\beqn}{\begin{eqnarray}}
\newcommand{\eeq}{\end{equation}}
\newcommand{\eeqn}{\end{eqnarray}}

\newcounter{mysection}

\begin{document}
\bibliographystyle{plain}
\pagestyle{plain}
\vspace*{1.5cm}
\begin{center} {\LARGE\bf
The finite one--dimensional\\ \rule{0mm}{1cm} wire problem}
\end{center}
\vskip2cm
\begin{center} {\large
Stefan~Kehrein${}^1$,
Christian~M{\"u}nkel${}^{2,}$\footnotemark[4]
and Kay~J.~Wiese${}^3$ }
\end{center}
\vskip1cm
\begin{center}
\begin{minipage}{13cm}
${}^1 \quad$\begin{minipage}[t]{12cm} \begin{flushleft}
Theoretische Physik III, Elektronische Korrelationen und Magnetismus,
Institut f\"ur Physik, Universit{\"a}t Augsburg,\\
86135~Augsburg, Germany  \end{flushleft}  \end{minipage} \\~\\
${}^2 \quad$\begin{minipage}[t]{12cm} \begin{flushleft}
Institut f\"ur Theoretische Physik, Universit\"at
Heidelberg,  \\
Philosophenweg 19, 69120~Heidelberg, Germany \end{flushleft} \end{minipage} 
\\ ~ \\
${}^3 \quad$\begin{minipage}[t]{12cm} \begin{flushleft}
FB Physik, Universit\"at GH Essen, 45117 Essen, Germany\end{flushleft} \end{minipage}
\end{minipage}
\end{center}
\footnotetext[4]{Present address: SAP AG, Neurottstra{\ss}e 16, 
69190 Walldorf, Germany}
\vfill\pagebreak
\begin{abstract} \noindent
We discuss an elementary problem in electrostatics: What does the charge  
distribution
look like for a free charge on a strictly one--dimensional wire of {\em  
finite} length? To the
best of our knowledge this question has so far not been discussed anywhere.  
One notices
that a solution of this problem is not as simple as it might appear  
at first sight.
\end{abstract}
\vfill\pagebreak
\section{Introduction}
Some time ago one of the authors of this paper was confronted with the following
question by a student: {\em What does the charge distribution look like for a free
charge on a strictly one--dimensional wire of finite length, if we consider  
the usual
Coulomb repulsion law?} Despite its rather trivial appearance this problem led to 
controversial discussions. The immediate ``obvious" answer by most
considering the problem for the first time is to suggest a charge  
distribution like the
one in Fig.~1: That is one intuitively expects an accumulation of charge
at the ends of the wire since there is no repelling charge outside. We urge  
the reader
to make up his mind too before reading on.

Let us remark that one--dimensional systems of electrons interacting with  
long--range
Coulomb forces have attracted much attention recently since it has become  
possible to
realize them experimentally as one--dimensional semiconductor  
structures~\footnotemark[3]. To avoid misunderstandings
we want to emphasize that this paper does not aim at contributing in this  
direction of research,
though this serves as an interesting background. Quantum effects and the fermionic
nature of electrons are not taken into account here. We rather want to introduce a
nice exercise in classical electrostatics for students or anybody else  
interested in
elementary problems. Only very elementary tools of mathematics and physics  
will be used,
still at first sight the answer might appear counter--intuitive and surprising. 

Let us make one more remark in order to avoid confusion. 
We do not discuss a problem in a one--dimensional world, but rather a 
one--dimensional problem defined in three--dimensional space. Therefore 
we use the three--dimensional Coulomb law $V(r)\propto |r|^{-1}$. Notice
that starting
from the one--dimensional Laplace equation the interaction potential has
the different form $V(r)\propto |r|$. This then defines the well--investigated
one--dimensional Coulomb gas model that has been solved exactly in the
thermodynamic limit independently by Prager~\footnotemark[4] and 
Lenard~\footnotemark[5]. Unfortunately the techniques used in these solutions
cannot be carried over to our problem~\footnotemark[6] and we have
to rely on other tools in the sequel. 

\section{Regularization prescriptions}
Once one starts analyzing the problem, one immediately notices that it
is ill--defined in its original formulation. For simplicity we will assume  
that the wire
has unit length throughout this work: The charge distribution~$\rho(x)$ is  
defined on
the interval~$[0,1]$ of the x--axis. In order to solve our problem we want
to minimize the energy functional
\beq  E[\rho]=\frac{1}{2}\:\int_0^1 dx \int_0^1 dy\;\;  
\frac{\rho(x)\,\rho(y)}{|x-y|}
\label{eq_energy}
\eeq
under the constraints of a fixed total charge~$Q$ here set to one
\beq  \int_0^1 dx\;\, \rho(x) = 1 \eeq
and
\beq  \rho(x)\geq 0 . \eeq
The integral (\ref{eq_energy}) diverges since
\beq \left( \int_0^{x-\eps} + \int_{x+\eps}^1 \right) \, dy\; \frac{1}{|x-y|} \;
\stackrel{\eps\rightarrow 0}{\leadsto} \; 2\ln\eps  .
\eeq
We will discuss two obvious possibilities to make the problem well--defined,  
i.e.~ways to
regularize the energy functional~(\ref{eq_energy}).
It is not immediately clear that both give the same answer.

\begin{itemize}
\item[(i)] Define the one--dimensional wire to be the limit of ellipsoids in  
three--dimensional
space when two semiprincipal axes of the ellipsoid shrink to zero. This way  
we use a
regularization by going to a well--defined problem in the embedding  
three--dimensional
space,  thereby avoiding the singular one--dimensional problem. This  
approach is discussed
in  \setcounter{mysection}{3} Sec.~\Roman{mysection}.
\item[(ii)] A physically appealing approach is to put $n$~equally charged classical
particles in equilibrium on the wire. Each particle has an individual  
charge~$q_i=1/n$. This
defines the looked for charge density in the limit~$n\rightarrow\infty$.  
Although this
regularization procedure requires more effort than~(i), it is more  
convincing in the sense of
being a ``microscopic" approach. This makes up the main part of our paper  
and is worked out in \setcounter{mysection}{4} Sec.~\Roman{mysection}.
\end{itemize}

\noindent
Of course other regularization procedures are also possible. A very natural  
choice would be
\beq
V_d(r)=\frac{1}{\sqrt{r^2+d^2}}\quad \stackrel{d\rightarrow  
0}{\longrightarrow} \quad
V(r)=\frac{1}{|r|}
\label{eq_transversereg}
\eeq
since this is usually used for one--dimensional systems interacting with  
long--range
Coulomb forces~\footnotemark[7]. The main advantage is that the  
one--dimensional Fourier
transform of $V_d(r)$ exists for finite~$d$. However, no exact solution
seems possible for finite~$d$ and it is therefore difficult to investigate
the limit~$d\rightarrow 0$.

\section{Shrinking ellipsoids}
Consider the conducting ellipsoid in Fig.~2 with semiprincipal axes~$a,b$
and~$c$. One can argue that for $b,c\rightarrow 0$ the ellipsoid shrinks
to a one--dimensional wire of finite length~$2a$.

Now the potential problem of a free charge~$Q$ on a conducting ellipsoid has been 
well--known for a long time and is treated in many advanced textbooks on  
electrostatics,
see for example Ref.~\footnotemark[1]. The explicit solution relies on the fact 
that the Laplace equation is separable in elliptic coordinates. Let us  
simply quote the result
for the potential
\beq
V(\xi)=\frac{Q}{8\pi}\:\int_\xi^\infty \frac{d\lambda}{\sqrt{ (\lambda+a^2)\,
(\lambda+b^2)\,(\lambda+c^2) } }
\label{eq_potential_ellipsoid}
\eeq
with $\xi=\xi(x,y,z)$ defined implicitly by
\beq
\frac{x^2}{a^2+\xi}+\frac{y^2}{b^2+\xi}+\frac{z^2}{c^2+\xi} =1 .
\eeq
The surface charge~$\sigma$ is given by the normal derivative of the  
potential at the surface
\beq
\sigma=-\left(\frac{\partial V}{\partial n} \right)_{\xi=0}  .
\eeq
In rectangular coordinates this is
\beq
\sigma=\frac{Q}{4\pi\,a\,b\,c}\:\frac{1}{\sqrt{
\displaystyle  \frac{x^2}{a^4}+\frac{y^2}{b^4}+\frac{z^2}{c^4} }} \ .
\label{eq_surfacecharge}
\eeq
For our purposes we can assume $b=c$. One would argue that
the total surface charge in the strip~$S$ in Fig.~2 collapses onto the line
charge~$\tau(x)\,dx$ of the wire at that point~$x$
\beq  \tau(x)=\sigma(x)\,\cdot\, S(x) ,
\label{eq_linecharge} \eeq
where $S(x)\,dx$ is the surface of the strip~$S$. It is a simple exercise in  
geometry to show
\beq
S(x)\,dx=2\pi\,b \sqrt{1-\frac{x^2}{a^2}\left(1-\frac{b^2}{a^2}\right) }\, dx .
\eeq
On the other hand, evaluating Eq.~(\ref{eq_surfacecharge}) at the surface gives
\beq
\sigma(x) = \frac{Q}{4\pi\,a\,b}\:\frac{1}{
\displaystyle\sqrt{ 1-\frac{x^2}{a^2}\left(1-\frac{b^2}{a^2}\right) }} \ .
\eeq
This implies that independent of~$b$ the line charge density defined like in 
Eq.~(\ref{eq_linecharge}) is constant along the wire
\beq  \tau(x)=\frac{Q}{2a} \ .
\label{eq_flat} \eeq
{\em According to this reasoning there are no finite size effects in our
one--dimensional wire problem: There is no accumulation of charge at the  
ends of the wire!}
\footnotemark[8]

The sceptical reader can certainly question the validity of this proof  by  
pointing out that
even for infinitesimal~$b$, in some respects the ellipsoid is in no way  
similar to a wire.
E.g.~the ends of the ellipsoid are always much thinner than its middle,
 and the smaller available space may compensate a charge accumulation
at the ends of the wire. Therefore we will
use a different, more microscopic regularization in the next section.

\section{Regularization by discretization}
We regularize the singular one--dimensional problem by considering $n$  
equally charged
classical particles with individual charges $1/n$ in the interval  
$\left[0,1\right]$. Due to
Coulomb repulsion there is a unique equilibrium state with charges at  
positions~$x_i$. In the
limit $n\rightarrow\infty$ this will define the continuum charge  
distribution on the wire that we
are interested in.

\subsection{Upper and lower bounds}
First of all we will show that the energy of the discretized problem  
diverges like $\ln n$ but
can still be determined within bounds of width $\half$. This will be done by  
calculating an
upper and a lower bound for the energy of the equilibrium distribution.

An upper bound is given by the energy $E_{\max}$ of a uniform distribution  
with charges
at positions $x_i=(i-1)/(n-1),~i=1\ldots n$
\beq
E_{\max}(n)\;\;=\;\;\frac{1}{2}\sum_{ {i,j=1}\atop{i\neq j} }^n
\frac{\displaystyle\frac{1}{n^2}}{|x_i-x_j|}
\;\;=\;\;\frac{n-1}{2n^2}\sum_{j=1}^n
  \left(
    \sum_{i=1}^{j-1} \frac{1}{i} + \sum_{i=1}^{n-j}\frac{1}{i}
  \right) .
\eeq
We use the well--known asymptotic behaviour (see e.g. formula~0.131 in
Ref.~\footnotemark[2])
\beq  \sum_{i=1}^n \frac{1}{i} = \ln n+\gamma+O(1/n) ,
\label{eq_asymptsum} \eeq
where $\gamma=0.5772\ldots$ is Euler's constant, and find
\beqn
E_{\max}(n)\;\;&=&\;\; \frac{n-1}{2n^2} \left( 2\sum_{j=1}^{n-1} \ln j +2n\,\gamma
+O(\ln n) \right) \nonumber \\
    \;\;&=&\;\;\ln n +\gamma -1 +O(1/n) \ln n .
\label{eq_upperbound}
\eeqn
In the last step we have used Stirling's formula.

In order to derive a lower bound $E_{\min}$ we first of all notice that the  
energy of a
discrete distribution can be split up as
\begin{equation}
	E=E_1+E_2+\ldots +E_{n-1}  ,
\end{equation}
where $E_k$ is the sum of energies of all particles to the $k$-th neighbours 
(see Fig.~3)
\beq E_k=\frac{1}{n^2} \sum_{ {i=1}\atop{i+k \leq n} }^n \frac{1}{|x_{i+k}-x_i|} . 
\label{eq_nextneighbour} \eeq
It is straightforward to see that each $E_k$ is bounded from below
by the energy obtained if a uniform distribution of the $k$-th neighbours
is assumed.
This gives e.g. in Eq.~(\ref{eq_nextneighbour})
\beq
  E_1 \ge \frac{1}{n^2} (n-1)^2 .
\eeq
Furthermore we have for even $n$
\beq
E_2 \ge \frac{2}{n^2} \left( \frac{n}{2}-1\right)^2
\eeq
and for odd $n$
\beq
  E_2 \ge \frac{1}{n^2} \left(
 \left(\frac{n+1}{2}-1\right)^2+\left(\frac{n-1}{2}-1\right)^2\right)
  \ge \frac{2}{n^2} \left(\frac{n}{2}-1\right)^2 .
\eeq
These results are easily generalized for arbitrary $k$
\begin{equation}
E_k \ge \frac{k}{n^2} \left( \frac{n}{k}-1\right)^2 .
\end{equation}
A lower energy bound $E_{\min}$ is therefore
\begin{eqnarray}
E_{\min} \;\;&=&\;\;
 \frac{1}{n^2}\sum_{k=1}^{n-1} k\left(\frac{n}{k}-1\right)^2
\;\;=\;\; \sum_{k=1}^{n-1} \left( \frac{1}{k}-\frac{2}{n}+\frac{k}{n^2} \right)
 \nonumber \\
    \;\;&=&\;\; \ln n +\gamma -\frac{3}{2} +O(1/n)
\label{eq_lowerbound}
\end{eqnarray}
where we have again used Eq.~(\ref{eq_asymptsum}).

Comparing Eq.~(\ref{eq_lowerbound}) with the upper bound from  
Eq.~(\ref{eq_upperbound}) one notices
that we have obtained rather strict limits for the energy of the equilibrium  
state with minimum
energy. Obviously a uniform distribution cannot be too far from the true  
equilibrium solution
with minimum energy. That this is indeed true will be shown in the next section.

\subsection{Rigorous results for the charge distribution}
{\large\bf Theorem~}{\em
Suppose the charge distribution $\rho_n(x)$ for $n$ particles converges for
$n \rightarrow \infty$ towards a continuous distribution $\rho(x)$
defined on $\left] 0,1 \right[ $. Then $\rho(x)$ is constant.
}
\vskip2mm
\noindent{\em Proof:}
Let us assume that the theorem is wrong and there exist $y_1,y_2 \in \left]  
0,1 \right[$ with
$\rho(y_1) > \rho(y_2)$, see Fig.~4. It shall be shown
that there exists an $N$ so that for $n>N$ transferring a particle
from $y_1$ to the middle of two particles at $y_2$ results in a
decrease of the energy. Hence the particles were not in equilibrium for $n>N$ and
the theorem is proven.

This can be seen as follows: First of all
we regard the energy needed to transfer a particle~$P$ from
$y_1$ to $y_2$ while neglecting the particles in
$\left[y_1-\varepsilon, y_1+\varepsilon \right]$ and in
$\left[y_2-\varepsilon, y_2+\varepsilon \right]$. Here $\varepsilon$ has been
chosen so small that both intervals are completely in $\left] 0,1 \right[$
and do not overlap. The change of energy is
\begin{equation} \label{25}
n\Delta E_I= \int \limits_0^1 dt \, \left(\frac{\tilde
 \rho(t)}{|y_1-t|}-\frac{\tilde \rho(t)}{|y_2-t|}\right)
 \  \stackrel{n\rightarrow \infty}{\longrightarrow} \ O(\E) \ ,
\end{equation}
where
\beq
\tilde\rho(t) = \left\{ \begin{array}{l} 0 \mbox{ in $\left[ y_1-\E,y_1+\E\right]$
 and
in $\left[ y_2-\E,y_2+\E \right] $} \\ \rho(t)  \mbox{ elsewhere.} \end{array}
 \right.
\eeq
It is essential to note that Eq.~(\ref{25}) does not contain
any term divergent in $n$, but has a finite and well-defined limit
for $n\to\infty$.
Let us now calculate the energy~$\Delta E_1$ of the particle~$P$ with  
respect to the other
particles in the interval $\left[ y_1-\E, y_1+\E \right]$. We can assume the  
distribution to be locally
uniform, deviations from that will only contribute in order~$\E^2$. Thus the  
particles in the
interval~$\left]y_1,y_1+\E\right]$ are at positions
\beq x_i=y_1+\frac{\E}{n q}\, i \ , \quad i=1\ldots n\,q  , \eeq
where $q=\E\,\rho(y_1)$ is the total charge in~$\left]y_1,y_1+\E\right]$.  
This gives
\begin{eqnarray}
n\,\Delta E_1 \;\;&=&\;\;\frac{2}{n}\:\sum_{i=1}^{n\,q} \frac{1}{|x_i-y_1|}
\;\;=\;\; 2 \sum_{i=1}^{n\,\E\,\rho(y_1)}\,\frac{\rho(y_1)}{i} +O(\E) \nonumber \\
\;\;&=&\;\; 2 \rho(y_1) \ln(\rho(y_1)\E n) +  O(n^0) \ ,
\end{eqnarray}
where the sum is done like in the previous section.
Particle~$P$ is placed in the middle between two particles at $y_2$ and this  
costs the
extra energy~$\Delta E_2$
\begin{eqnarray}
  n \Delta E_2 \;\;&=&\;\; 2  
\sum_{i=1}^{n\,\E\,\rho(y_2)}\,\frac{\rho(y_2)}{i-\half}+O(\E)
 \nonumber \\
  \;\;&=&\;\; 2 \rho(y_2) \ln(\rho(y_2)\E n) +  O(n^0) \ .
\end{eqnarray}
The total change of energy is
\begin{equation}
  n(\Delta E_I +\Delta E_1 -\Delta E_2) =  2\left( \rho(y_1)-\rho(y_2)  
\right) \ln n
 + O(n^0) \ .
\end{equation}
For $n$ large the expression is dominated by the first term on the r.h.s., i.e.\
the charge distribution is stable only for $\rho(y_1)=\rho(y_2)$.
\hfill \raisebox{-0.5mm}[0mm][0mm]{$\Box$}

\vskip1mm
\noindent Two remarks are to be made:
\begin{itemize}
\item
The proof does not say anything about the endpoints of the interval. It can  
be shown
that if the charge density at the  endpoints is well--defined in the  
continuum limit, it will obey
the inequality $\rho(0) \le 2 \rho(0.5)$.
\item
The proof can be generalized for all potentials
 $V(x-y)=|x-y|^{-\alpha}$ with $\alpha \ge 1$. Therefore only potentials  
with longer range
forces than the Coulomb potential (i.e. potentials with~$\alpha<1$) can show finite
size effects in the wire problem.
\end{itemize}

\subsection{Some estimates for the discrete charge distribution}
After proving that the continuum charge distribution is flat, let us go back to the
discretized version of the problem. We want to derive some estimates for the 
distribution of the charges close to the ends of the wire. In combination
with numerical results in the next subsection, this will help us to  
reconcile the somehow
counter--intuitive picture of a flat continuum charge distribution.

We use the same notation as in \setcounter{mysection}{4} Sec.~\Roman{mysection}.1. 
The distances between particle~$i$
and particle~$i+1$ are denoted by $d_i=x_{i+1}-x_i$. Then force equilibrium for the
second, third etc. particle on the wire means that the following set of  
equations is
fulfilled
\begin{eqnarray}
\frac{1}{d_1^2}&=&\frac{1}{d_2^2}+\frac{1}{(d_2+d_3)^2}
+\frac{1}{(d_2+d_3+d_4)^2 }+\ldots \nonumber \\
\frac{1}{(d_1+d_2)^2}+\frac{1}{d_2^2}&=&\frac{1}{d_3^2}+\frac{1}{(d_3+d_4)^2}+
\frac{1}{(d_3+d_4+d_5)^2}+\ldots
\label{eq_equilibrium} \\
\frac{1}{(d_1+d_2+d_3)^2}+\frac{1}{(d_2+d_3)^2}+\frac{1}{d_3^2}&=&
\frac{1}{d_4^2}+\frac{1}{(d_4+d_5)^2}+\frac{1}{(d_4+d_5+d_6)^2}+\ldots \nonumber
\end{eqnarray}
The first and the $n$--th particle are trivially at positions $x_1=0$ and $x_n=1$ 
and no equilibrium condition can be formulated for them. We sum the first~$l$ of
the above equations ($l\le n/2$) and subtract equal terms on both sides
\beq
\sum_{m=1}^l \frac{1}{\displaystyle \left(\sum_{k=1}^m d_k\right)^2 }
= \sum_{m=1}^l \sum_{a=1}^m
\frac{1}{\displaystyle \left( \sum_{b=1}^m d_{l+1+b-a} \right)^2}
+\sum_{m=l+1}^{n-2} \sum_{a=1}^{l}
\frac{1}{\left(\displaystyle \sum_{b=1}^m d_{a+b}\right)^2 } ~.
\eeq
For simplicity we assume that $n$ is even. Then obviously the largest distance
between particles is $d_{max}=d_{n/2}$ and we have the following inequality
\beq
\sum_{m=1}^l \frac{1}{\displaystyle \left(\sum_{k=1}^m d_k\right)^2 }
\geq \frac{1}{d_{max}^2} \left( \sum_{m=1}^l \frac{1}{m}
+l \sum_{m=l+1}^{n-2} \frac{1}{m^2} \right) .
\eeq
The first sum is like Eq.~(\ref{eq_asymptsum}). The second sum can be  
evaluated using the integral approximation 
\beq
l \sum_{m=l+1}^{n-2}\frac1{m^2} \approx l \int\limits_{l+1/2}^{n-3/2}
\mbox d m\, \frac1{m^2} = \frac l{l+1/2}-\frac l{n-3/2} \ .
\eeq
One then finds
\beq
\sum_{m=1}^l \frac{1}{\displaystyle \left(\sum_{k=1}^m d_k\right)^2 }
\geq \frac{1}{d_{max}^2} \left( \ln l +\gamma + 1- \frac{l}{n} \right)
\eeq
for large values of $l$, $n$.
For $l=n/2$ one has $d_1<d_2<\ldots <d_{n/2}$,
therefore
\beq
\sum_{m=1}^{n/2} \frac{1}{\displaystyle \left(\sum_{k=1}^m d_k\right)^2 }
\leq \frac{1}{d_1^2} \sum_{m=1}^{n/2} \frac{1}{m^2}
\leq \frac{1}{d_1^2} \; \frac{\pi^2}{6} ~.
\eeq
We finally get
\beq \label{eq_estimate1}
d_1^2 \leq d_{max}^2 \frac{ \displaystyle \frac{\pi^2}{6} }{ \displaystyle
\ln n -\ln 2 +\gamma +\frac{1}{2} } .
\eeq
Lower bounds can be given too using the first equilibrium condition in
Eq.~(\ref{eq_equilibrium})
\beqn
\frac{1}{d_1^2}&=&\frac{1}{d_2^2}+\frac{1}{(d_2+d_3)^2}
+\frac{1}{(d_2+d_3+d_4)^2 }+\ldots \nonumber \\
&\leq& \frac{1}{d_2^2} \sum_{m=1}^{n-2} \frac{1}{m^2}
\quad\leq\quad \frac{1}{d_2^2} \frac{\pi^2}{6}  \\
\Rightarrow \qquad d_1^2 &\geq& \frac{6}{\pi^2} d_2^2 ~ .
\label{eq_estimate2}
\eeqn
Eq.~(\ref{eq_estimate1}) says that for large $n$, the distance of the first
two particles on the wire is arbitrarily smaller than the distance of
two particles in the middle of the wire. Numerical calculations in the next
subsection will show that the equality is nearly realized in
Eq.~(\ref{eq_estimate2}). Thus the particles close to the ends of the wire  
show strong
finite size effects. However, these effects vanish in the continuum limit
as we have proved in the last subsection. This discrepancy is resolved by
noticing that the finite size region observed in the discretized problem
shrinks for~$n\rightarrow\infty$ as will be demonstrated in the next 
subsection.

\subsection{Numerical results}
In order to gain a better understanding of the problem, we have also
used numerical methods to find the equilibrium position of the $n$ particles
sitting on the line of unit length. We have employed the Hybrid 
Monte--Carlo~\footnotemark[9] algorithm
that updates the positions of the individual particles
until they find their equilibrium positions:
The molecular dynamics part of the algorithm
moves all particles according to the electrostatic
forces, which are computed only once for all
particles at each step. The Monte--Carlo update
scheme ensures that the algorithm is exact,
i.e. that equilibrium is reached. During the
simulation, the ``temperature'' was reduced
as the charges moved more and more closely
to their equilibrium positions. 
Convergence was
verified by checking that the ground state energy did not decrease any more
for additional sweeps within a given numerical precision. In general, 
convergence was very quick since the starting point of the simulation
with all particles equidistant is quite ``close'' to the equilibrium position
for the reasons explained in Sec.~IV.2. Simulations with up to $n=8193$ particles 
have been performed. 

The first interesting quantity investigated numerically 
is the ground state energy $E(n)$
as a function of the number of particles~$n$ plotted in Fig.~5. Also drawn
are the upper and lower bounds derived in Eqs.~(\ref{eq_upperbound})
and (\ref{eq_lowerbound}). One notices that the measured energies seem
to settle on the upper bound with good precision for $n\rightarrow\infty$.
This raises the interesting (and so far unanswered) question whether the
upper bound~(\ref{eq_upperbound}) becomes the asymptotically exact result
for the ground state energy. 

Let us
also emphasize that the data in Fig.~5 paedagogically demonstrate the danger
of extrapolating to $n\rightarrow\infty$ based on simulations for
finite~$n$, at least for systems with long range forces: Extrapolating on 
the basis of the data for $n<1000$ particles the upper bound would
eventually be violated (see the curve in Fig.~5)! {\em Only for more than
1000~particles the curve for $E(n)$ bends over and the extrapolation 
respects the exact results~\footnotemark[10].}
 
In Fig.~6 the quotient of the smallest (at one end of the wire) to the
largest (in the middle of the wire) particle distance is plotted as a
function of~$n$. The upper bound from Eq.~(\ref{eq_estimate1}) is
respected, in particular the interparticle distance at the end of the
wire becomes much smaller than the maximum distance. On the other hand,
the ratio of the distance of the first to the second particle~$d_1$ and
the second to the third particle~$d_2$ approaches a constant nonzero value 
for $n\rightarrow\infty$. This is shown in Fig.~7 together with the lower bound
derived in Eq.~(\ref{eq_estimate2}). One clearly sees the strong
finite size effects already mentioned in the previous subsection.

However, when one looks at the distribution of all charges along the
wire plotted in Fig.~8 for various values of~$n$, one notices that
the regions with strong finite size effects close to the ends of the
wire shrink very slowly with increasing~$n$, probably as 
slow as $1/\sqrt{\ln(n)}$, as suggested by Eq.~(\ref{eq_estimate1}). 
One can interpret these
numerical results as follows: Eventually the continuum
charge distribution appears flat, although for a given~$n$ the first
few particles do never approach equidistant positions. But in the
continuum limit $n\rightarrow\infty$ these non--equidistant regions 
eventually vanish as compared to the rest of the wire. {\em This scenario
combines the analytical and numerical results established in this
work.}

\section{Summary}
We have discussed the seemingly trivial, at least conceptually simple
problem of a free charge on a one--dimensional
wire of finite length. The first thing to notice was that this problem is  
ill--defined
in its original formulation due to the diverging ground state energy. The  
question how the
continuum charge distribution looks like can only be answered after  
introducing some
regularization procedure.

Two particularly intuitive regularizations have been employed in this paper;  
shrinking
ellipsoids in \setcounter{mysection}{3} Sec.~\Roman{mysection}
and charge discretization in
\setcounter{mysection}{4} Sec.~\Roman{mysection}. Both led to the answer
that the continuum charge distribution is flat, that is {\em there are no  
finite size effects!}
The interested reader will be able to find other regularization  
prescriptions that lead to
the same answer.

A flat charge distribution will also be found for potentials
\beq V(x-y)\propto |x-y|^{-\alpha} \eeq
with shorter range forces than the Coulomb potential, that is generally for 
$\alpha\geq 1$. For
exponents $\alpha<1$ the problem is well--defined and finite without the need for
regularization. It is easy to show that then there are finite size effects.  
Thus the
Coulomb law is the limiting case between finite size effects and no finite size
effects in a strictly one--dimensional wire of finite length, see Fig.~9.

One interesting problem left open in this respect is the analytic form of  
the continuum
charge distribution for exponents $\alpha <1$. Another question that we have
not been able to answer is whether the ground state energy as a function of the
number of particles~$n$ in Sec.~\Roman{mysection} is really given by the  
upper bound
Eq.~(\ref{eq_upperbound})  plus subdominant corrections~$O(\ln n/n)$.
The numerical results in Fig.~5 seem to indicate this.

\section*{Acknowledgments}
The authors are grateful to the unknown student who has provided us with this
little riddle. We also want to thank a large number of anonymous colleagues for
drawing pictures like Fig.~1 when we annoyed them with our problem. We acknowledge
useful discussions with F.~Wegner.
\vfill

{~ }
\footnotetext[1]{
J.A. Stratton,
{\it Electromagnetic Theory},
(Mc Graw--Hill, New York and London, 1941), 1st ed., pp.~207--210.}

\footnotetext[2]{
I.S. Gradshteyn and I.M. Ryzhik,
{\it Table of Integrals, Series and Products},
(Academic Press, San Diego, 1994), 5th ed.}

\footnotetext[3]{
A.R.~Go\~{n}i, A.~Pinczuk, J.S.~Weiner, J.M.~Calleja, B.S.~Dennis, L.N.~Pfeiffer
and K.W.~West, ``One--Dimensional Plasmon Dispersion and Dispersionless  
Intersubband
Excitations in GaAs~Quantum Wires",
Phys. Rev. Lett. {\bf 67}, 3298--3301 (1991).}

\footnotetext[4]{
S.~Prager, ``The One--Dimensional Plasma'', 
in {\it Advances in Chemical Physics}, edited by I.~Prigogine
(Interscience Publishers, New York, 1962), Vol.~IV.}

\footnotetext[5]{
A.~Lenard, ``Exact Statistical Mechanics of a One--Dimensional System with
Coulomb Forces'', J.~Math. Phys.~{\bf 2}, 682--693 (1961).}

\footnotetext[6]{
A general criterion for the exact solvability of one--dimensional
problems with some interaction potential has been given in 
R.~J.~Baxter, ``Many--body Functions of a One--Dimensional Gas'',
Phys. Fluids~{\bf 7}, 38--43 (1964). This criterion is e.g. fulfilled
for the Coulomb gas model, but one can check that it does not
hold for the interaction potential discussed in this paper.}

\footnotetext[7]{
A.~Gold and A.~Ghazali, ``Analytical results for semiconductor quantum--well wires:
Plasmons, shallow impurity states, and mobility",
Phys. Rev.~B{\bf 41}, 7626--7640 (1990).}

\footnotetext[8]{One can also derive Eq.~(\ref{eq_flat}) directly from  
Eq.~(\ref{eq_surfacecharge}) by performing the limit $b,c\rightarrow 0$  
there.}

\footnotetext[9]{S.~Duane, A.~D.~Kennedy, B.~J.~Pendleton and D.~Roweth,
``Hybrid Monte--Carlo'', Phys. Lett.~B{\bf 195}, 216--222 (1987);
see also
B.~Mehlig, D.~W.~Heermann and B.~M.~Forrest,
``Hybrid Monte--Carlo Method for Condensed--Matter Systems'', 
Phys. Rev.~B{\bf 45}, 679--685 (1992).}

\footnotetext[10]{The observation that an exact bound would be violated
based on our simulations for $n<1000$~particles was the motivation
for pushing the simulations to $n=8193$. This also demonstrates the
dangers of using numerical simulations without analytical ``guidelines''.}

\pagebreak
\setcounter{section}{1}
\noindent
\subsection*{Figure captions}
\vskip1cm
\renewcommand{\labelenumi}{{Fig.} \arabic{enumi}.}
\begin{enumerate}
\item A frequent first guess for the charge distribution on a finite wire of  
unit length.
Some readers might also suggest divergencies at the ends of the wire.
\item A free charge on a conducting ellipsoid.
\item Sum of energies to the $k$-th neighbours.
\item Proof by contradiction --- a nonflat charge distribution cannot be stable in 
the limit~$n\rightarrow\infty$.
\item Ground state energy $E(n)$ for a system of $n$ particles with  
individual charges $1/n$
interacting with Coulomb potentials on a wire of unit length.
\item The quotient of smallest to largest particle distance as a function of  
the number
of particles~$n$.
\item The quotient of the distances of particle \#1 and \#2 and of particle  
\#2 and \#3 on
the wire as a function of~$n$.
\item Charge distribution on the wire for various numbers of particles~$n$.
$x_i$ is the position of particle~\#$i$. Nearest neighbor
distances $x_{i+1}-x_i$ are plotted as a function of the position along
the wire.
\item Finite size effects on one--dimensional wires of finite length for  
potentials\linebreak
$V(x-y)\propto |x-y|^{-\alpha}$.

\end{enumerate}

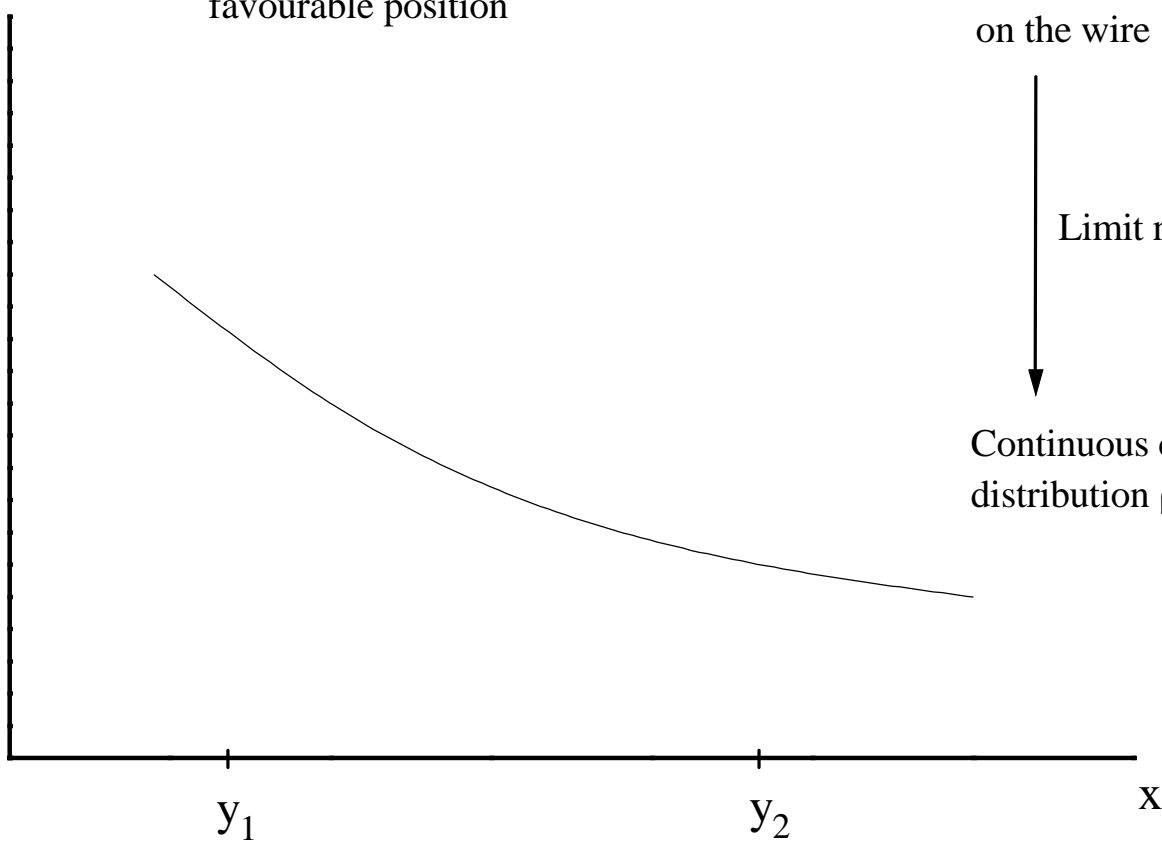
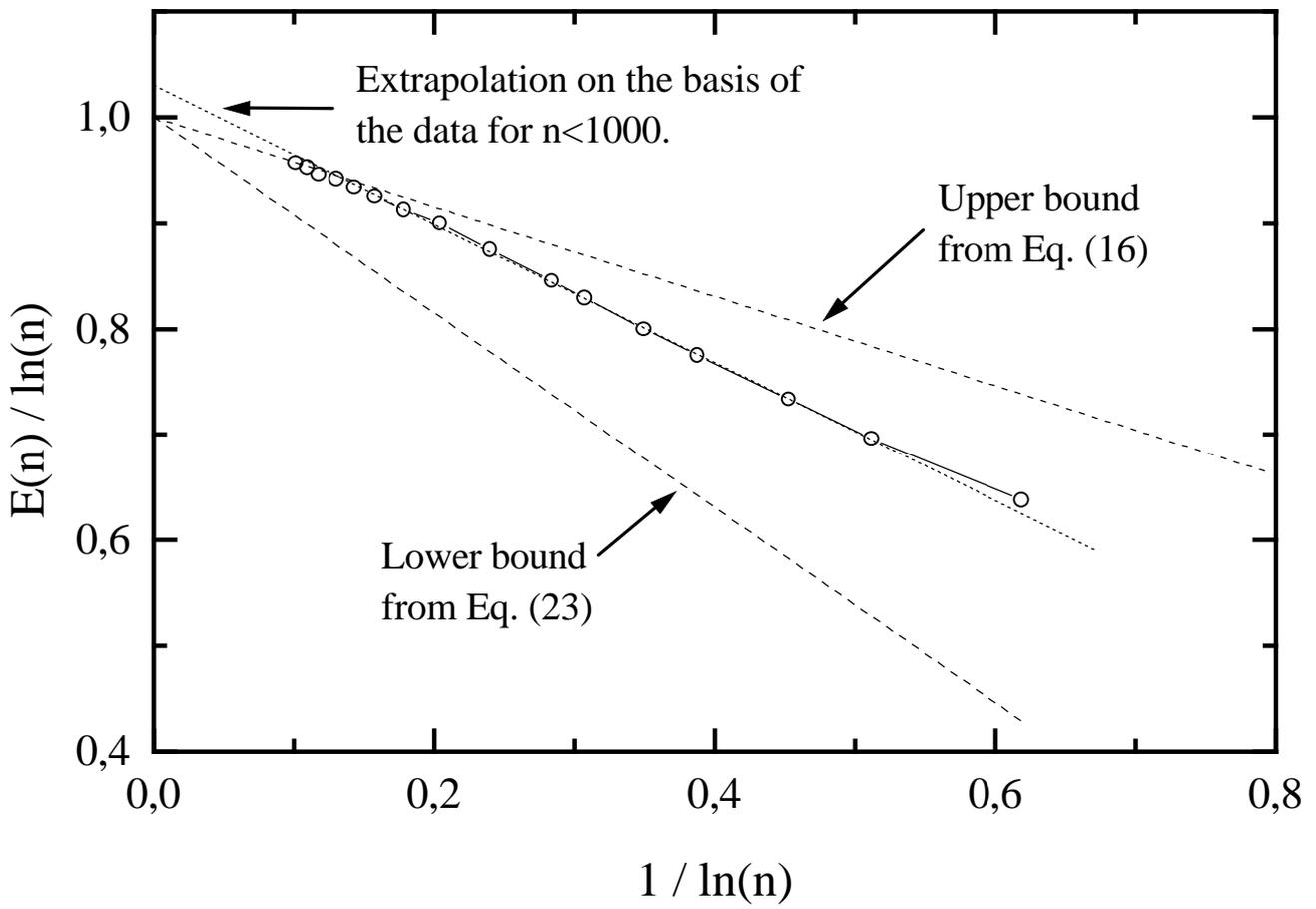
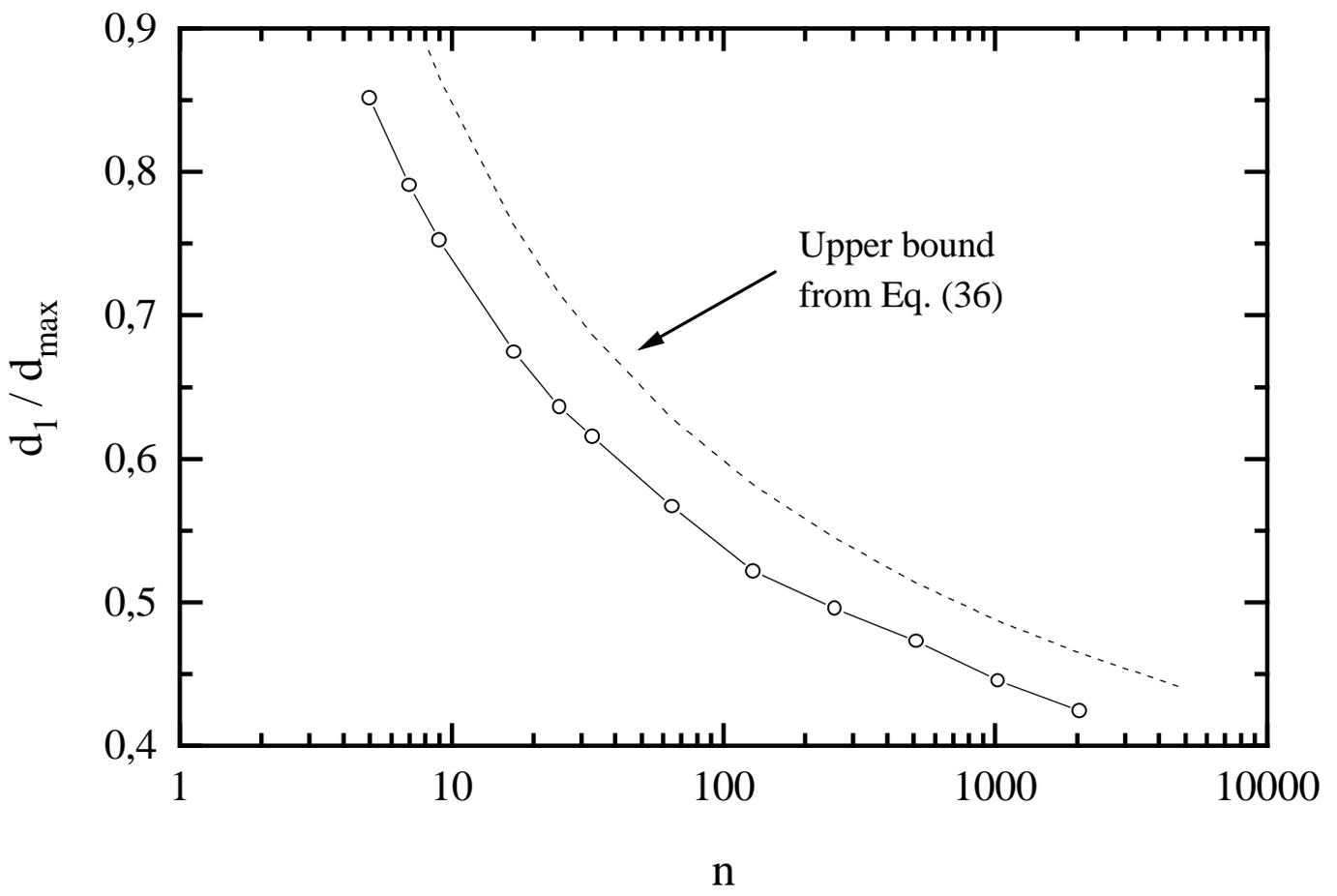
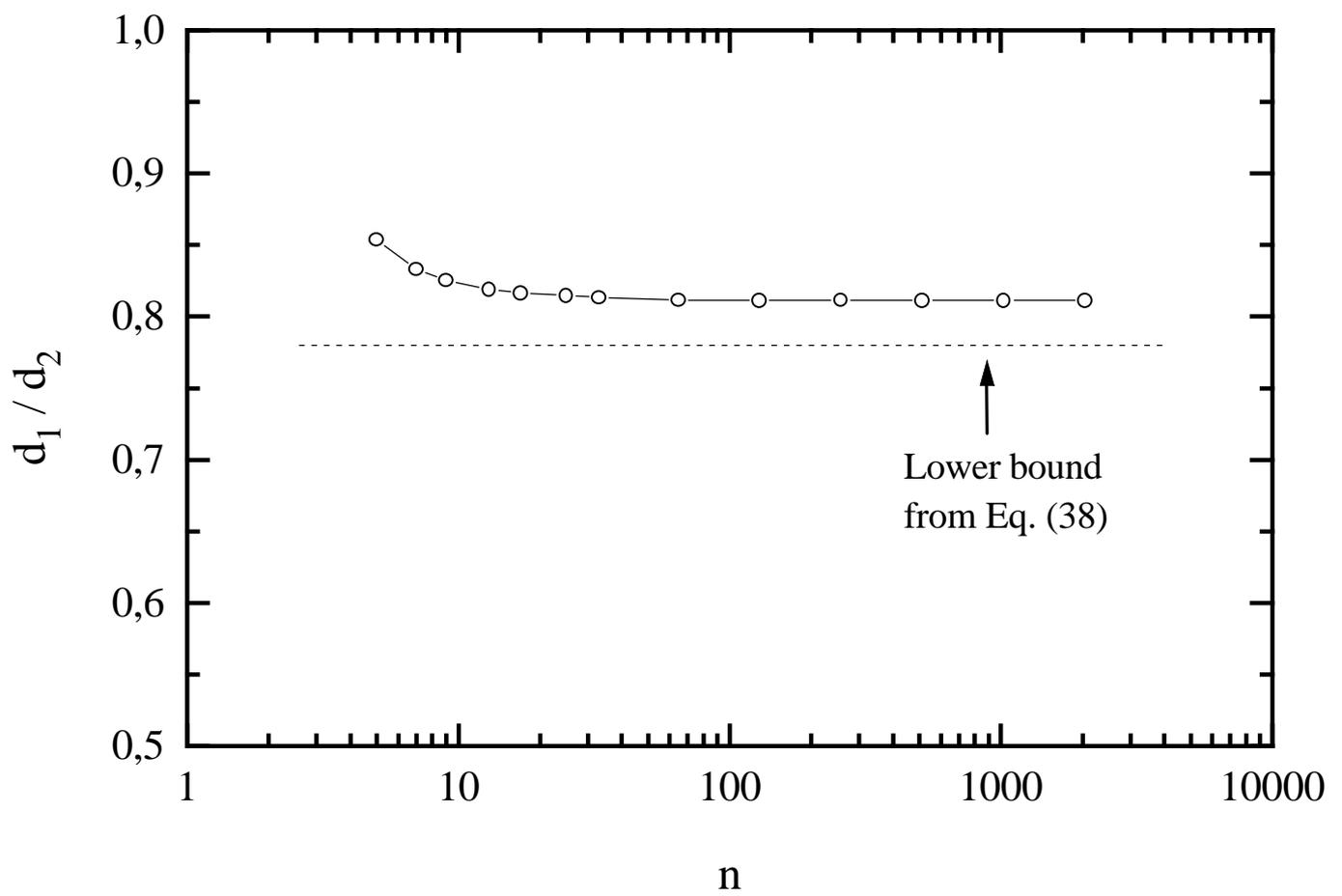
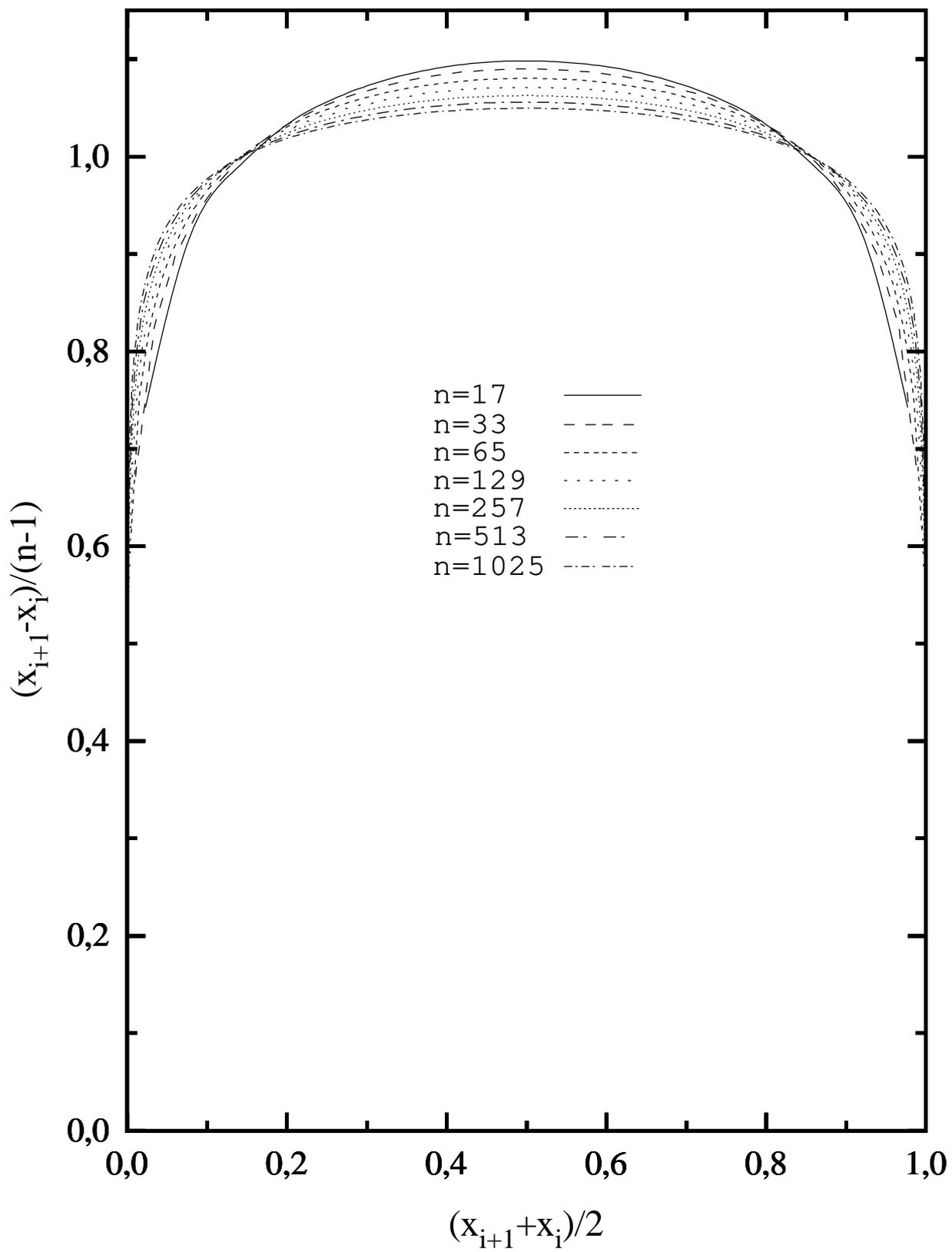
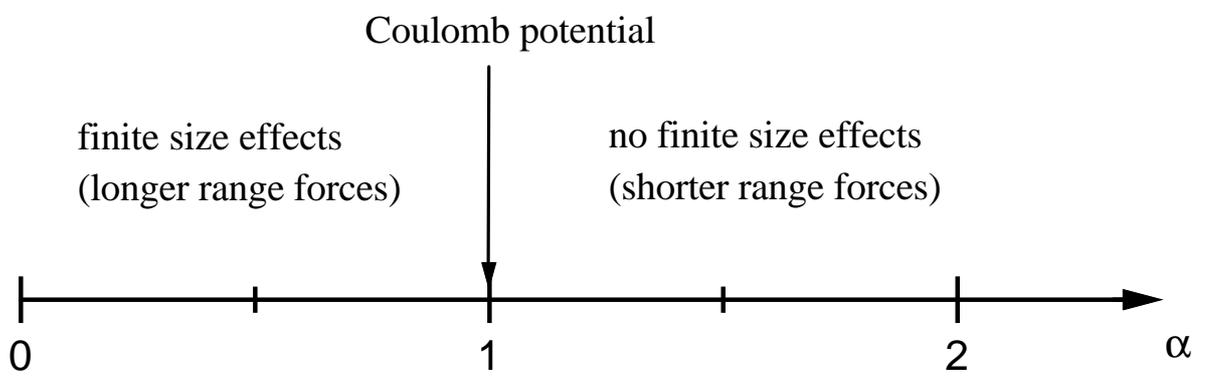
\end{document}